\documentclass[preprintnumbers,amsmath,amssymb,nofootinbib,groupedaddress,superscriptaddress]{revtex4}

\usepackage{graphicx}
\usepackage{dcolumn}
\usepackage{epsfig}
\usepackage[dvips]{color}
\usepackage{bm}
\usepackage{fancybox}

\usepackage{mathrsfs} % \mathscr

\usepackage{booktabs}

\usepackage{framed} % \begin{shaded}
\definecolor{shadecolor}{gray}{0.90}

\allowdisplaybreaks[4] % page break

\newcommand{\eps}{\varepsilon}

\newcommand{\bi}{\begin{itemize}}
\newcommand{\ei}{\end{itemize}}

\newcommand{\be}{\begin{equation}}
\newcommand{\ee}{\end{equation}}
\newcommand{\bea}{\begin{eqnarray}}
\newcommand{\eea}{\end{eqnarray}}

\newcommand{\ie}{{\it i.e.}}

\newcommand{\cf}{{\it cf.}}

\newcommand{\eq}{Eq.}

\newcommand{\fig}{Fig.}

\newcommand{\Ref}{Ref.}

\newcommand{\equ}[1]{\eq~(\ref{equ:#1})}
\newcommand{\figu}[1]{\fig~\ref{fig:#1}}

\begin{document}

%%%%%%%%%%%%%%%%%%%%%%%%%%%%%%%%%%%%%%%%%%%%%%%%%%%%%%%%%%%%%%%%%%%%%%

\title{\LARGE \vspace*{2cm} Summary report of MINSIS workshop in Madrid\\}
\vspace*{0.3cm}

\author{Rodrigo Alonso}\affiliation{Univ. Autonoma de Madrid and IFT-UAM/CSIC, Spain}
\author{Stefan Antusch}\affiliation{Max-Planck Institute, Munich, Germany}
\author{Mattias Blennow}\affiliation{Max-Planck Institute, Munich, Germany}
\author{ Pilar Coloma}\affiliation{Univ. Autonoma de Madrid and IFT-UAM/CSIC, Spain} 
\author{Andre de Gouvea}\affiliation{Northwestern University, USA}
\author{Enrique Fernandez-Martinez}\affiliation{Max-Planck Institute, Munich, Germany}
\author{Belen Gavela}\affiliation{Univ. Autonoma de Madrid and IFT-UAM/CSIC, Spain}
\author{Concha Gonzalez-Garcia}\affiliation{ICREA, Univ. de Barcelona, Spain and Stony Brook, USA} 
\author{Sergio Hortner}\affiliation{Univ. Autonoma de Madrid and IFT-UAM/CSIC, Spain} 
\author{Marco Laveder}\affiliation{Padua Univ., Italy}
\author{Tracey Li}\affiliation{Durham Univ., United Kingdom} 
\author{Jacobo Lopez-Pavon}\affiliation{Univ. Autonoma de Madrid and IFT-UAM/CSIC, Spain}
\author{Michele Maltoni}\affiliation{Univ. Autonoma de Madrid and IFT-UAM/CSIC, Spain}
\author{Olga Mena}\affiliation{IFIC, Univ. de Valencia/CSIC, Spain}
\author{Pasquale Migliozzi}\affiliation{INFN, Napoli, Italy and CERN, Switzerland}
\author{Toshihiko Ota}\affiliation{Max-Planck Institute, Munich, Germany} 
\author{Sergio Palomares Ruiz}\affiliation{ CFTP, Instituto Superior Tcnico, Lisbon, Portugal} 
\author{Adam  Para}\affiliation{Fermi National Accelerator Laboratory, USA}
\author{Stephen Parke}\affiliation{Fermi National Accelerator Laboratory, USA}
\author{Nuria Rius}\affiliation{IFIC, Univ. de Valencia/CSIC, Spain}
\author{Thomas Schwetz-Mangold}\affiliation{Max-Planck-Institut fuer Kernphysik, Heidelberg, Germany} 
\author{F.~J.~P. Soler}\affiliation{Glasgow Univ., United Kingdom} 
\author{Michel Sorel}\affiliation{IFIC, Univ. de Valencia/CSIC, Spain} 
\author{Osamu Yasuda}\affiliation{Tokyo Metropolitan Univ., Japan}  
\author{Walter Winter}\affiliation{Wuerzburg Univ., Germany}

\vspace*{0.3cm}

\date{\today}



%%%%%%%%%%%%%%%%%%%%%%%%%%%%%%%%%%%%%%%%%%%%%%%%%%%%%%%%%%%%%%%%%%%%%%
\begin{abstract}
\vspace*{1cm}
Recent developments on tau detection technologies 
and the construction of high intensity neutrino beams 
open the possibility of a high precision search for 
non-standard $\mu$ - $\tau$  flavour transition with neutrinos 
at short distances.  
The MINSIS --- Main Injector Non-Standard Interaction Search-- 
is a proposal under discussion to realize such precision measurement.
This document contains the proceedings of the workshop which took place
on 10-11 December 2009 in Madrid to discuss both the physics reach as well 
as the experimental requirements for this proposal\footnote{ 
The original slides can be found at the workshop 
web-site~\cite{MINSIS2009Madrid}}.
\end{abstract}

\maketitle

\pagebreak 
%%%%% contents %%%%%
\tableofcontents

\pagebreak 

%%%%% main part %%%%%
\section{Overview of talks}

%%%%%%%%%%%%%%%%%%%%%%%%%%%%%%%%%%%%%%%%%%%%%%%%%%%%%%%%%%%%%%%%%%%%%%
\subsection{MINSIS --- Main Injector Non-Standard Interaction Search
      --- {\it A. Para}}
%--------------------------------------------------------------------%
\label{para}
Neutrino oscillations are the only known, so far, examples of the lepton
flavor violating processes. \ The interactions responsible for the
oscillations-induced lepton flavor violation are the same as the ones
responsible for the neutrino masses and they are likely to operate at very
high energies, similar to the GUT scale. On the other hand it is quite
possible that there are other processes, beyond the minimal standard model,
which operate at much lower energies and which induce lepton flavor
violations. The examples of mechanisms include:

\begin{itemize}
\item  Non-unitarity of the neutrino mixing matrix: TeV scale see-saw,
inverse see-saw could be a typical example

\item  Leptoquarks mediated interactions

\item  R-parity SUSY particles violating interactions

\item  Charged Higgs mediated interactions
\end{itemize}

Neutrino oscillations constitute an irreducible background for such possible
rare processes, hence the search for the non-standard interactions ought to
be conducted at the short baseline, where oscillations probability is known
to be very small. Muon neutrino to tau neutrino appears to be particularly
promising proposition. The existing limit provided by the short-baseline
neutrino oscillations experiments NOMAD and CHORUS is of the order of $%
10^{-4}$.

The next generation of the short-baseline tau appearance experiments with
the sensitivity improved by up to two orders of magnitude was well
developed, but abandoned shortly after the discovery of the neutrino
oscillations by the SuperK experiment and a determination that the
atmospheric neutrino oscillations are induced by the mass difference of 
the order 
of % --- or : corrected by ota 
$3\times 10^{-3}~{\rm eV}^{2}$. This discovery has demonstrated that the
oscillation probability is extremely low at the short-baseline oscillations.

What was a fatal blow for the neutrino oscillations search is a great
advantage for the search for non-standard neutrino oscillations: signatures
for potential new physics may appear on top the extremely low or no
background. The proposals for the next generation short baseline oscillation
search offered a major improvement of the sensitivity event with the
detector technology of the late 90's. Such a search is complementary to the
direct searches conducted at the LHC and it has somewhat different
sensitivities. Advances in the experimental techniques in conjunction with
modern high intensity neutrino beams offer an attractive prospect of
discovery of the non-standard neutrino interactions or a significant
improvement of the limits on the strength of such interactions:

\begin{itemize}
\item  high intensity neutrino beam with flexible energy has been
constructed at Fermilab

\item  a near detector hall, housing the near MINOS detector and the MINERvA
experiment, has been constructed and it offers sufficient floor space for
new additional experiment

\item  $\tau $ neutrino interactions have been observed using a new concept
of the Emulsion Cloud Chamber (ECC)

\item  OPERA, a very large $\tau $ appearance experiment using the ECC
technology has been constructed and it is currently operating in Gran Sasso

\item  automatic scanning and measurement of the emulsion films have been
developed for the OPERA experiment, making it feasible to analyze vary large
detector volumes 
\end{itemize}

These initial studies indicate that the beam related backgrounds (primarily
a $\tau $ neutrino component of the beam) as well as the instrumental
background (primarily induced by mis-identified charm events) can be
controlled at the levels allowing the  tau appearance experiment at the
short distance in the NUMI beam with a  sensitivity better than $10^{-6}$.
Detailed optimization of the beam and detector configuration is necessary to
establish a credible case, though. A significant body of experience
developed by the OPERA experiment is of the key importance for such an
effort.

%%%%%%%%%%%%%%%%%%%%%%%%%%%%%%%%%%%%%%%%%%%%%%%%%%%%%%%%%%%%%%%%%%%%%%
\subsection{$\nu_{\tau}$ detection:
       the CHORUS and OPERA experiences --- {\it P. Migliozzi} }
%--------------------------------------------------------------------%

\label{migliozzi}
The critical points in designing an experiment that aims at improving
the sensitivity on $\theta_{\mu\tau}$ of the CHORUS and NOMAD
experiments are the understanding of the background and its
suppression.  Here in the following we briefly summarize the
experience gathered from the emulsion based experiments CHORUS and
OPERA and possible improvements.

The CHORUS experiment was designed to search for $\nu_\mu \rightarrow
\nu_\tau$ oscillations through the observation of charged-current
interactions $\tau \rightarrow \nu_\tau + X$ followed by the decay of
the $\tau$ lepton, directly observed in a nuclear emulsion target.  As
discussed in detail in Ref. \cite{Eskut:2007rn}, the main source of
background in the CHORUS experiment originates from the poor
efficiency in measuring the momentum and the charge of the decay
products. Indeed, the excellent sensitivity of the nuclear emulsions
in detecting also nuclear recoils (which is extremely important in
rejecting hadron reinteractions mimicking a decay topology) is spoiled
by the low efficiency in performing a kinematical analysis of the
events. The achieved background level (normalized to charged-current
interactions) by the CHORUS Collaboration is $\thickapprox10^{-3}$.

An evolution of the ``pure'' nuclear emulsion target concept is the so
called Emulsion Cloud Chamber (ECC). It consists of photographic
emulsion films interleaved with lead (or iron) plates, where the
emulsion films act as micrometric tracking device and the lead plates as
passive material (target). The main advantages of 
% the : corrected by ota
this approach, on top of a micrometric decay topology reconstruction,
are: momentum measurement (with an accuracy of about 20\%) of charged
particles by exploiting their Multiple Coulomb Scattering on lead;
electron/pion separation and gamma identification through the
electromagnetic shower reconstruction.

The ECC technique has been adopted by the OPERA Collaboration to search
for $\nu_\mu \rightarrow \nu_\tau$ oscillations on the CNGS beam
\cite{Acquafredda:2009zz}. The synergy among the micrometric decay
topology reconstruction and an efficient kinematical analysis allowed
the OPERA Collaboration to estimate a background level  of
$\thickapprox10^{-5}$. The breakdown of the background, normalized to
the charged-current events, for each decay channel is shown in 
Tab.~\ref{Tab:BG-ECC}.

\begin{table}
\begin{center}
\begin{tabular}{|c|c|c|c|}
  \hline
                               & $\tau\rightarrow e$ & $\tau\rightarrow \mu$ & $\tau\rightarrow h$  \\
                                 \hline
  Charm background             & $7\times10^{-6}$    & $3\times10^{-7}$      & $5\times10^{-6}$  \\
    \hline
  Large angle $\mu$ scattering &                     & $4\times10^{-6}$      &   \\
    \hline
  Hadronic background          &                     & $3\times10^{-6}$      & $4\times10^{-6}$  \\
    \hline
  Total=$2\times10^{-5}$       & $7\times10^{-6}$    & $7\times10^{-6}$      & $9\times10^{-6}$  \\
  \hline
\end{tabular}
\end{center}
\caption{Background estimation of Emulsion Cloud Chamber by 
the OPERA collaboration.}
\label{Tab:BG-ECC}
\end{table}

From Tab.~\ref{Tab:BG-ECC}, it is evident that even if an OPERA-like
experiment restricts the search to the muonic channel the background
is still well above the $\sim10^{-6}$ level. Indeed there are two
sources of background that dominate: the large angle muon scattering
and the hadron reinteractions. The knowledge of the former background
is limited by the absence of data and by the fact that GEANT
simulation does not take into account nuclear form factors. The
present estimate is based on upper limits from measurements performed
in the past, while a calculation that accounts for nuclear form
factors gives a background five times smaller. On the other hand, the
hadron reinteraction background in an ECC is amplified with respect to
a detector that exploits a pure emulsion target since it is not
possible to detect nuclear fragments produced in the
interaction. Indeed, in OPERA this background is suppressed by
applying strong kinematical cuts.

In order to efficiently exploit also the electronic and hadronic
channels it is of the outmost importance to measure the charge of the
daughter particles. This would allow a reduction of the background to
a level comparable to the one of the muonic channel.

\vspace{0.2cm}
What can be the ultimate background level? --- It is difficult to say
without a better understanding of the large angle muon scattering
background and of an optimization of the kinematical analysis. The
latter strongly depends on the availability of a magnetic
field. Indeed, if an ECC is immersed into a magnetic field, only the
muonic channel is studied and the large angle muon scattering is
confirmed to be smaller as expected from numerical calculations, then
a background level of $\sim10^{-6}$ can be achieved.

The feasibility of a background level $\sim10^{-7}$ is a real
challenge and deserves dedicated studies both on the detector and
analysis optimization. Such a studies are more challenging and
mandatory if one wants to search for decay by exploiting all $\tau$
decay channels.

%%%%%%%%%%%%%%%%%%%%%%%%%%%%%%%%%%%%%%%%%%%%%%%%%%%%%%%%%%%%%%%%%%%%%%
\subsection{NSI for Fermilab to DUSEL --- {\it S. Parke}}
%--------------------------------------------------------------------%
\label{parke}
Fermilab is planning a new neutrino beamline to send a conventional
neutrino superbeam to DUSEL, the new underground laboratory
at Homestake, South Daykota.  Initially the beamline will be powered 
by the 700 kW of protons from the Fermilab Main Injector but will
be upgraded to more than 2 MW once the new proton source, Project X, 
is completed. The detector complex at DUSEL will consists
of up to 300 ktons of water Cerenkov (WC) detectors and up to 50 ktons 
of Liquid Argon TPCs (LAr) in some mix, e.g. 2 $\times$ 100 kton modules 
of WC and a 17 kton module of LAr. 
The combination of Project X and the detectors at DUSEL will have 
a sensitivity to $\sin^2 2\theta_{13} \approx 0.001$ and be able to 
determine the hierarchy and measure the CP violating phase with 
reasonable accuracy down to $\sin^2 2\theta_{13} \approx 0.01$.  
Preliminary studies indicate that with this facility one could new sets 
limits on Non-Standard Neutrino Interactions especially for 
$\epsilon_{e\mu}$, $\epsilon_{e\tau}$ and $\epsilon_{\mu \tau}$
at better than 0.1 level.

%%%%%%%%%%%%%%%%%%%%%%%%%%%%%%%%%%%%%%%%%%%%%%%%%%%%%%%%%%%%%%%%%%%%%%
 \subsection{ Non-unitarity PMNS matrix (Theory)      --- {\it S. Antusch}}
%--------------------------------------------------------------------%
\label{antusch}
Non-unitarity of the leptonic mixing matrix is a typical signal of new
physics. Intuitively, non-unitarity results when the light neutrinos
of the Standard Model (SM) mix with heavier states, for instance with
heavy fermionic singlet states with masses above the energies of a
given experiment. While the full mixing matrix is unitary, the
effective mixing matrix relevant for the low energy experiment is just
a submatrix and it is in general non-unitary \cite{Langacker:1988ur}.

Non-unitarity and neutrino masses can be introduced in an effective
theory approach by adding only two gauge invariant operators to the
SM. A minimal possibility, referred to as Minimal Unitarity Violation
(MUV) \cite{Antusch:2006vwa} consists in adding the lepton number violating dimension 5
(Weinberg) operator for neutrino masses and the lepton number
conserving dimension 6 operator which contributes to the kinetic
energy term of the neutrinos (but not of the charged leptons). {\em
The bounds on non-unitarity and the relevance for Minsis are
summarized by M. Blennow.}

The non-unitarity effects can be sizable, for instance, when neutrino
masses are generated in a SM extension by fermionic singlets at
comparably low energies, i.e.\ close to the electroweak scale. The
smallness of neutrino masses can be accommodated in this scheme by an
approximately conserved lepton number symmetry. As a consequence, the
dimension 6 operator effects generically dominate over the effects
from the lepton number violating dimension 5 operator.

This can also have interesting consequences for the thermal
leptogenesis mechanism, and the relation between non-unitarity and
leptogenesis has briefly been discussed \cite{Antusch:2009gn}: On the one hand, the
flavoured decay asymmetries for leptogenesis can be strongly enhanced,
and on the other hand additional flavour-equilibrating interactions in
the thermal bath can become important. Both effects are due to the
dimension 6 operator which induces non-unitarity.

%%%%%%%%%%%%%%%%%%%%%%%%%%%%%%%%%%%%%%%%%%%%%%%%%%%%%%%%%%%%%%%%%%%%%%
 \subsection{Non-unitarity PMNS matrix (Bound) --- {\it M. Blennow}}  
%--------------------------------------------------------------------%
\label{blennow}
The current non-oscillation bounds on non-unitarity \emph{(see text by
  Stefan Antusch)} are derived from the measurements of $W$ and $Z$
decay widths, rare lepton decays such as $\mu \to e \gamma$, and
universality tests of weak interactions (see
\cite{Antusch:2006vwa,Antusch:2008tz}). The bound put on the
$\varepsilon_{\mu\tau}$ parameter, which is the most relevant for the
MINSIS experiment, is $|\varepsilon_{\mu\tau}| < 5\cdot10^{-3}$ at
90~\%~CL. However, if it is assumed that the non-unitarity is due to
mixing with some heavy states, then it is required that $\varepsilon$
is a negative semi-definite matrix and the relation
$|\varepsilon_{\mu\tau}|^2 <
|\varepsilon_{\mu\mu}\varepsilon_{\tau\tau}|$ then imposes the
stronger bound of $|\varepsilon_{\mu\tau}| < 1.1\cdot10^{-3}$.

\begin{figure}[tb]
  \begin{center}
    \includegraphics[width=0.5\textwidth]{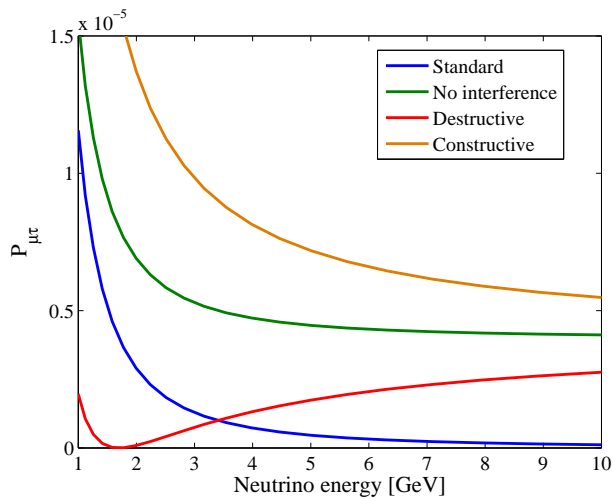}
    \caption{Oscillation probabilities for $|\varepsilon_{\mu\tau}| =
      10^{-3}$ at $L=1$~km depending on whether interference is
      constructive, destructive, or absent. The standard oscillation
      probability is shown for comparison.}\label{near-interference}
  \end{center}
\end{figure}

To leading order in $\varepsilon_{\mu\tau}$ and $L$, the oscillation
probability at the MINSIS near detector is given by
\begin{equation}
  P_{\mu\tau} \simeq |2\varepsilon_{\mu\tau}-iH_{\mu\tau}L|^2 = 
  4|\varepsilon_{\mu\tau}|^2 + |H_{\mu\tau}L|^2 
  + 4\ {\rm Im}(\varepsilon_{\mu\tau}^* H_{\mu\tau})L,
\end{equation}
where $H_{\mu\tau} \simeq \Delta m_{31}^2 \sin(2\theta_{23})/(4E)$. If
we just regard the non-unitiarity term, then MINSIS sensitivity to
$|\varepsilon_{\mu\tau}|$ would be $0.5\sqrt{P_{\mu\tau}~{\rm
    sensitivity}}$. This would mean that the bound could be
strengthened by an order of magnitude if a $P_{\mu\tau}$ sensitivity
of $10^{-7}$ could be achieved. Although the standard oscillation term
is most likely just beyond the reach of the MINSIS sensitivity, the
interference term (see, \emph{e.g.}, \cite{FernandezMartinez:2007ms})
could be sizable if $\varepsilon_{\mu\tau}$ is large, see
Fig.~\ref{near-interference}. This effect would in principle allow
for the detection of CP-violation in the non-standard sector if enough
precision could be obtained. It should also be noted that this effect
is not unique to non-unitarity, but would also be present in more
general scenarios of non-standard neutrino interactions.

%%%%%%%%%%%%%%%%%%%%%%%%%%%%%%%%%%%%%%%%%%%%%%%%%%%%%%%%%%%%%%%%%%%%%%
\subsection{ Non-standard neutrino Interactions (Theory) 
  --- {\it  E. Fern\'{a}ndez-Mart\'{i}nez}}
%--------------------------------------------------------------------%
\label{enrique}

Neutrino non-standard interactions (NSI) are a very widespread and convenient way of parametrizing the effects of new physics in neutrino oscillations experiments. NSI can affect the neutrino production and propagation processes, as well as the neutrino propagation through matter. For the MINSIS experiment the relevant NSI are those contributing to neutrino production and detection via hadronic interactions, these NSI can be described by effective four-fermion operators of the form:
\begin{equation}
 \mathcal L^{\rm NSI} = -2\sqrt{2} G_F \eps^{ud}_{\alpha\beta}  
 [\bar u \gamma^\mu P d][\bar\ell_\alpha \gamma_\mu P_L \nu_\beta] + {\rm h.c.}.
\label{eq:NSI}
\end{equation}
Notice that different chirallity structures other than the example of Eq.~(\ref{eq:NSI}) can be considered. While vector or axial couplings have the same chirallity structure than the Standard Model (SM) contribution and can therefore interfere with them, for the neutrino production via pion decay NSI involving scalar or pseudoscalar couplings are enhanced with respect to the SM model contribution since a chirallity flip of the charged lepton is not required for spin conservation.

The model independent bound on the NSI relevant for MINSIS that can be set by the phenomenological implications of the operator of Eq.~(\ref{eq:NSI}) is $\eps^{ud}_{\mu \tau}<1.8 \cdot 10^{-2}$ at the $90 \%$ CL~\cite{Biggio:2009nt}. However, saturating the bound in particular extensions of the SM is rather challenging, since NSI are expected to be produced together with charged fermion non-standard interactions due to SU(2) gauge invariance~\cite{Antusch:2008tz,Gavela:2008ra}. The implications of the gauge invariant counterpart of the operators relevant for MINSIS involving charged fermions are discussed in the talk by T.~Ota and~\cite{Antusch:2010fe}, here we will instead explore the possibility of extensions of the SM that induce neutrino NSI but do not generate the related charged fermion interactions.

At $d=6$ this can be realized in two ways. The first realization involves the addition of fermion singlets (right-handed neutrinos) to the SM particle content. The light active neutrinos mix with the extra singlets inducing a non-unitary mixing matrix that modifies the neutrino couplings to the $W$ and $Z$ and that generates NSI upon integrating out the gauge bosons. However, as was described in the talk by M.~Blennow, the bounds on non-unitarity are stronger, in particular, for the MINSIS experiment $\eps_{\mu \tau}<1.1 \cdot 10^{-3}$ \cite{Antusch:2006vwa,Antusch:2008tz}. The second possibility involves the addition of singly charged scalar SU(2) singlets that couple to a pair of lepton doublets. Upon integrating out the singlets the following operator is induced:
\begin{eqnarray}\label{Eq:AntisymmDim6}
(\overline{L^c}_\alpha i \sigma_2 L_\beta) (\bar L_\gamma i \sigma_2 L^c_\delta) 
\;.
\end{eqnarray} 
This operator can contribute to matter NSI of $\nu_\mu$ and $\nu_\tau$ with electrons, but these NSI are not relevant for the MINSIS experiment, as matter effects are too weak for the short baseline considered.

At the level of $d=8$ operators the only extensions of the SM inducing NSI but avoiding their charged fermion counterparts involve either one of the two realizations at $d=6$ (and the consequent strong bounds), or rather unnatural fine-tunings between different mediators to cancel the undesired charged fermion contribution, both at the tree and the loop level~\cite{Antusch:2008tz,Gavela:2008ra,Biggio:2009kv}. 

The most promising NSI realization avoiding the bounds set by gauge invariance thus seems non-unitarity. However, a sensitivity to the oscillation probability of $10^{-7}$ at the $90 \%$ CL would be necessary to improve the present bounds.

%%%%%%%%%%%%%%%%%%%%%%%%%%%%%%%%%%%%%%%%%%%%%%%%%%%%%%%%%%%%%%%%%%%%%%
 \subsection{ Non-standard neutrino Interactions (Bound)
     ---  {\it T. Ota}}
%--------------------------------------------------------------------%
\label{ota}

After integrating out the heavy degrees of freedom, 
a model would be described by the Standard Model (SM) interactions
and non-renormalizable interactions of the SM fields, 
which respect the SM gauge symmetries.
Non-Standard neutrino Interactions (NSI) may emerge as such effective 
interactions at the electroweak scale.
On theoretical aspects of NSIs, 
see the talk presented by E. Fernandez-Martinez.
In general, 
NSIs are constrained by the corresponding charged Lepton Flavour Violating 
(LFV) processes through the SM gauge symmetries.
An important exception is the Minimal Unitarity Violation (MUV) 
in which NSIs are induced 
with the (spontaneous) violation of the SM gauge symmetries,
and therefore, the NSIs are not directly constrained by the 
charged LFV (for a theoretical motivation of MUV and constraints, 
see the talks given by S. Antusch and M. Blennow).
In this talk, we investigated the bounds to the coefficients of the 
four-Fermi NSIs which were relevant to MINSIS  
from the various LFV rare tau decay processes. 
We found that the coefficients were constrained 
at $\mathcal{O}(10^{-4}) \times G_{F}$ where $G_{F}$ was Fermi constant. 
With this value, the ratio of the signal and the standard model process
is naively expected to be $\mathcal{O}(10^{-7}\text{-}10^{-8})$
which is far below the scope of the expected MINSIS sensitivity.
However, a scalar-mediated NSI which is described with the 
effective Lagrangian
\begin{align}
\mathcal{L}_{\text{eff}} 
=&
\frac{G_{F}}{\sqrt{2}}
\varepsilon
[\bar{\nu}_{\tau} (1+\gamma^{5}) \mu]
[\bar{d} \gamma^{5} u]
\label{eq:Leff-chiralenhance}
\end{align}
can make an enhanced effect in a pion decay 
(see e.g., Ref.~\cite{Herczeg:1995kd}).
The decay rate calculated from the Lagrangian 
Eq.~\eqref{eq:Leff-chiralenhance} is 
\begin{align}
\Gamma (\pi^{+} \rightarrow \mu^{+} \nu_{\tau})
=
\left|
\varepsilon \omega_{\mu}
\right|^{2}
\cdot
\Gamma (\pi^{+} \rightarrow \mu^{+} \nu_{\mu}),
\end{align} 
where $\omega_{\mu}$ is the chiral enhancement factor which is 
about a factor of ten. 
With this enhancement factor, the ratio can be expected to 
become $7.9 \cdot 10^{-5}$ which is expected to be 
achieved in MINSIS experiment~\cite{Antusch:2010fe}.
The bounds are summarized in Tab.~II.

\begin{table}[thb]
\begin{tabular}{cccc}
%%%%%%%%%%%%%%%%%%%%%%%%%%%%%%%%%%%%%%%%%%%%%%%%%%
\hline \hline
%%%%%%%%%%%%%%%%%%%%%%%%%%%%%%%%%%%%%%%%%%%%%%%%%%
Beam (channel) & $2L2Q$ & $4L$ & $NU$
\\
%%%%%%%%%%%%%%%%%%%%%%%%%%%%%%%%%%%%%%%%%%%%%%%%%%
\hline
%%%%%%%%%%%%%%%%%%%%%%%%%%%%%%%%%%%%%%%%%%%%%%%%%%
$\pi (\mu \rightarrow \tau)$ 
& $7.9\cdot 10^{-5}$ 
& n/a
& $4.4 \cdot 10^{-6}$
\\
%%%%%%%%%%%%%%%%%%%%%%%%%%%%%%%%%%%%%%%%%%%%%%%%%%
$\beta (e \rightarrow \tau)$
&
$<10^{-6}$
&
n/a
&
$1.0\cdot 10^{-5}$
\\
%%%%%%%%%%%%%%%%%%%%%%%%%%%%%%%%%%%%%%%%%%%%%%%%%%
$\mu (\mu \rightarrow \tau)$
&
$<10^{-6}$
&
$1.0\cdot 10^{-3}~(3.2 \cdot 10^{-5})$
&
$4.4 \cdot 10^{-6}$
\\
%%%%%%%%%%%%%%%%%%%%%%%%%%%%%%%%%%%%%%%%%%%%%%%%%%
$\mu (e \rightarrow \tau)$
&
$<10^{-6}$
&
$1.0 \cdot 10^{-3}~(3.2 \cdot 10^{-5})$
&
$1.0 \cdot 10^{-5}$
\\
%%%%%%%%%%%%%%%%%%%%%%%%%%%%%%%%%%%%%%%%%%%%%%%%%%
\hline \hline
\end{tabular}
\caption{Bounds at the 90~\% CL on the probability of
 tau appearance at a near detector in a neutrino beam
 from $\pi$ decay, $\beta$ decay or $\mu$ decays for 
 the three types of new physics, 
 NSI with two leptons and two quarks ($2L2Q$),
 NSI with four lepton doublets ($4L$),
 and the non-standard effect induced from non-unitarity 
 of the PMNS matirx ($NU$). 
 Two different values, the bound to the effective four-Fermi
 interactions and that under the assumption of the singlet scalar
 mediation
(in parenthesis), are shown in the column of leptonic NSI. 
 Table taken from Ref.~\cite{Antusch:2010fe}.
}
\end{table}

%%%%%%%%%%%%%%%%%%%%%%%%%%%%%%%%%%%%%%%%%%%%%%%%%%%%%%%%%%%%%%%%%%%%%%
\subsection{Minimal Flavour violation at MINSIS
    ---   {\it R. Alonso}}
%--------------------------------------------------------------------%
\label{alonso}
The aim of our work was to confront the parameters in our simplest
minimal feour violation (MFV) model with the bounds on non-unitarity
and determine if the sensitivity expected for MINSIS would give us
further information.

Our model is the simplest MFV type I seesaw\footnote{%
Minimal Flavour Seesaw Models~\cite{Gavela:2009cd}.
%M.B. Gavela, (Madrid, Autonoma U. & Madrid, IFT) , 
%T. Hambye, (Brussels U.) , 
%D. Hernandez, (Madrid, Autonoma U. & Madrid, IFT) , 
%P. Hernandez, (Valencia U. & Valencia U., IFIC) .
%Published in JHEP 0909:038,2009.
%e-Print: arXiv:0906.1461 [hep-ph]
}. 
Simplest, as there is only one fermionic singlet $N$ and the Dirac-Mass
connected $\overline{N}'$ added to the Standard Model that possess a
global U(1) symmetry. Discussion of such models was carried by Wyler,
Wolfenstein, Mohapatra, Valle $\ldots$ It also occurs that the number of
parameters in our model almost equals the 
%the : corrected by ota
low energy range ones, so we are able to express the formers in terms of
mixing angles 
%an : correted by ota
and phases.

The result of the comparison was that a \textbf{sensitivity of
$10^{-7}$ would improve the restrictions} coming from current
non-unitarity data. It would even \textbf{improve the results of
experiments such as MEG} (with it's huge expected sensitivity)
\textbf{for a certain region in our parameter space}. That is so
because there our model predicts a cancellation of the $\nu_e$
coupling in the yukawas and therefore the interesting place to look is
a $\nu_\tau$-$\nu_\mu$ transition experiment such as MINSIS.

\begin{figure}[tb]
  \begin{center}
\includegraphics[height=2.75in]{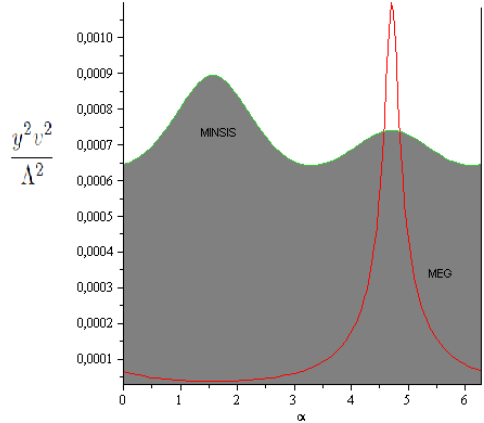}
\end{center}
\vspace{-0.8cm}
\caption{Allowed parameter region in the scenario of 
Minimal Flavour Violation.}
\label{Fig:Alonso}
\end{figure}

Here we plot the allowed values for a quotient of parameters in our
       model as a function of the Majorana phase $\alpha$
       (Fig.~\ref{Fig:Alonso}). We can see how MINSIS improves the MEG
       bound for an interval in $\alpha$ when $\theta_{13}=0.19$.

%%%%%%%%%%%%%%%%%%%%%%%%%%%%%%%%%%%%%%%%%%%%%%%%%%%%%%%%%%%%%%%%%%%%%%
\subsection{NSI bounds from ICECUBE
    ---   {\it O. Mena}}
%--------------------------------------------------------------------%
\label{mena}
We focus here on nonstandard interactions which may affect the neutrino
propagation in matter.
Constraining the new parameters of
the effective low energy theory which parametrizes the new physics
responsible for the nonstandard neutrino interactions might need (a) very
intense and (b) high energy ($\mathcal{O}$ 10 GeV) neutrino beams.
Far future neutrino facilities, as neutrino factories and/or superbeams
provide an ideal setup to test neutrino-matter nonstandard interactions,
see for instance Ref.~\cite{Kopp:2008ds} and references therein.

Cosmic ray interactions in the atmosphere give a natural beam of
neutrinos, with a steeply falling spectrum which covers several orders
of magnitude in energy (from hundreds of MeV to hundreds of TeV), with
a peak around 1~GeV. Atmospheric neutrinos in the GeV range have been
used by the Super-Kamiokande detector (SK) to provide evidence for
neutrino oscillations.  Straight up atmospheric neutrinos traverse the
Earth's core and are extremely sensitive to neutrino matter
interactions. Consequently, atmospheric neutrinos at GeV energies may
also constitute an ideal tool to test neutrino nonstandard
interactions~\cite{Bergmann:1999pk,GonzalezGarcia:1998hj,Fornengo:2001pm,GonzalezGarcia:2004wg},
see Ref.~\cite{Friedland:2004ah} for an analysis using the SK phase I
data set.

The Icecube Deep Core Subarray (ICDC)~\cite{deepcore,deepcore2},
a low energy extension of the IceCube detector, will accumulate an enormous
number of muon events from atmospheric muon neutrino charged current 
interactions, down to muon energies as low as 5 GeV. 
These muon events are usually considered as a background astrophysical 
neutrino searches.  In this
% paper : corrected by ota
talk we will show that these `` muon neutrino background events''
provide a great opportunity for measuring neutrino-matter nonstandard
interactions.  We concentrate on $\epsilon_{e\tau}$, $\epsilon_{\mu
\tau}$ and $\epsilon_{\tau \tau}$, setting the remaining nonstandard
interaction parameters to zero.  Furthermore, we will assume that
$\epsilon_{e\tau}$
%, : corrected by ota
and $\epsilon_{\mu\tau}$ are real. Our results are similar to those
obtained in Ref.~\cite{Friedland:2004ah} using the SK phase I data
set~\cite{inprep}.  However, further improvements in the sensitivity
to the neutrino-matter NSI parameters can be done if the $\nu_\tau$
identification in the Deep Core detector becomes feasible.

%%%%%%%%%%%%%%%%%%%%%%%%%%%%%%%%%%%%%%%%%%%%%%%%%%%%%%%%%%%%%%%%%%%%%%
\subsection{Probing the Seesaw Scale --- 
       from nano to mega electron-volts 
    ---   {\it A. de Gouv\^{e}a}}
%--------------------------------------------------------------------%
\label{gouvea}

Sterile neutrinos are among the simplest and most benign extensions of
the standard model of particle physics. They are gauge singlet
fermions and can only couple to the standard model at a renormalizable
level through a Yukawa interaction (${\cal L}_{\rm int}=-y(LH)N$,
where $L$ are the lepton doublets, $H$ is the Higgs scalar doublet and
$N$ are the gauge singlet fermion fields, aka right-handed neutrinos,
aka sterile neutrinos. $y$ are the Yukawa couplings). The most general
Lagrangian consistent with the standard model augmented by $n$ gauge
singlet fermions $N$ contains the Yukawa couplings above and Majorana
masses for the $N$ fields: ${\cal L}_{\rm mass}=-(M/2) NN$. After
electroweak symmetry breaking, this Lagrangian describes $3+n$
electrically neutral, generically massive fermions and can fit all
experimental data if one judiciously chooses the values of $y$ and $M$
(and $n$).

For any value of $M$, there is an associated value of $y$ that allows
one to fit all the neutrino oscillation data, at least
superficially. For example, in the case $M\equiv0$, the neutrinos are
massive Dirac fermions and their mass matrix is given by $m_D=yv$,
where $v$ is the Higgs boson vacuum expectation value. In this case,
the data require $y\sim 10^{-11}$ and the Lagrangian has an exact
global, nonanomalous $U(1)_{B-L}$ symmetry. This symmetry is broken
when both $y$ and $M$ are nonvanishing, a fact that teaches us that
any value of $M$ is technically natural (as defined by 'tHooft).

For non-zero values of $M$, this so-called ``Seesaw Lagrangian''
predicts that the neutrinos are Majorana fermions, and that there
exists more neutrinos than the three active ones that have already
been accounted for. If $M\lesssim 1$~MeV, these extra neutrinos are
light sterile neutrinos that can only be directly probed by
neutrino-related experiments, including searches for neutrino
oscillations at short baselines (most relevant for MINSIS), searches
for kinematical effects of neutrino masses, especially studies of the
end-point of the spectrum of beta-rays, and searches for neutrinoless
double-beta decay. In the case of neutrinoless double-beta decay, the
prediction is that, in spite of the fact that there are more neutrinos
and that all are Majorana fermions, the expected rate vanishes with 
very good precision as long as all $M$ values are smaller than about an MeV. 

\begin{figure}[tb]
\begin{center}
\includegraphics[width=0.6\textwidth]{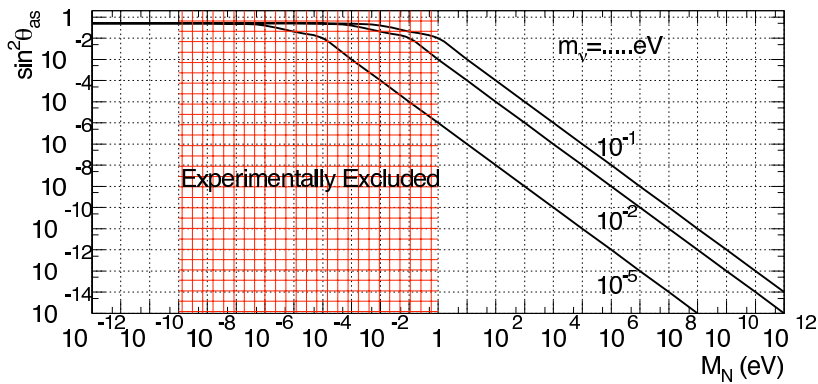}
\end{center}
\caption{The mixing angle $\theta_{as}$ 
between active and sterile neutrinos
as a function of the right-handed neutrino mass $M_{N}$
for different values of the mostly active neutrino masses 
$m_{\nu}$~\cite{deGouvea:2009fp}.}
\label{Fig:deGouvea}
\end{figure}

In the case of searches for neutrino oscillations, one can relate
       values of $M$ and the ``active'' neutrino oscillation
       parameters to the sterile--active mixing angles to which the
       different oscillation probabilities are sensitive. This means
       that once new, heavier neutrino mass eigenstates are
       discovered, along with their flavor composition, it will be
       possible, in principle, to test whether the seesaw Lagrangian
       describes the new degrees of freedom and to measure the
       Lagrangian parameters. Conversely, if no new neutrino mass
       eigenstates are discovered, one might be able to rule a range
       of potential values of the seesaw energy scale (i.e., values of
       $M$ will be ruled out). A summary of what one can already say
       about $M$, and what the expected values of the active--sterile
       mixing angles are is depicted Fig.~\ref{Fig:deGouvea}.
       Searches for $\nu_{\mu}\to\nu_{\tau}$ at short baselines are
       sensitive to $\theta_{\mu s}^2\theta_{\tau s}^2$.  The
       figure~\ref{Fig:deGouvea} hints that an experiment sensitive to
       $P_{\mu\tau}\gtrsim 10^{-6}$ should be sensitive to $M\lesssim
       100~{\rm eV}$. The exact capability of MINSIS and 
other short-baseline neutrino oscillation experiments to detect 
seesaw neutrinos is still to be ascertained.

%%%%%%%%%%%%%%%%%%%%%%%%%%%%%%%%%%%%%%%%%%%%%%%%%%%%%%%%%%%%%%%%%%%%%%
\subsection{Sterile neutrino mixings and near detectors
    ---   {\it O. Yasuda}}
%--------------------------------------------------------------------%
\label{yasuda}       
Sensitivity of a neutrino factory, with a near detector
or with far detectors, to sterile neutrino mixings
was reviewed, based on the two works \cite{Donini:2001xy,Donini:2008wz}.
In the case of a neutrino factory with a near detector
at the oscillation maximum $\Delta m^2_{41}L/4E\simeq\pi/2$
(with far detectors for $\Delta m^2_{41}\gtrsim$0.1eV$^2$),
we can improve the sensitivity to $4|U_{e4}U_{\mu4}|^2$,
$4|U_{e4}U_{\tau4}|^2$ and $4|U_{\mu4}U_{\tau4}|^2$
by 2, 4, 1 orders (by 1.5, 2.5, -1 orders)
of magnitude compared to the present bound,
respectively (See Figs. 2, 4 and 5 in Ref.\,\cite{Yasuda:2010rj}).
Near $\tau$ detectors are useful not only to improve
sensitivity to sterile neutrino mixings by themselves,
but also to reduce the systematic errors of the far $\tau$ detectors.

%%%%%%%%%%%%%%%%%%%%%%%%%%%%%%%%%%%%%%%%%%%%%%%%%%%%%%%%%%%%%%%%%%%%%%
\subsection{Sterile neutrinos in the MINSIS experiment
    ---   {\it T. Li} 
       {\it and}
       {\it J. L\'{o}pez-Pav\'{o}n}}
%--------------------------------------------------------------------%
\label{li}
We have performed a preliminary analysis of the sensitivity of the
MINSIS experiment to sterile neutrinos, using the $3+1$ model as an
initial study. We have assumed a NO$\nu$A beam (peaking at $\sim$ 2
GeV) and use a flux and detector mass of 4 kton; the total number of
charged-current $\nu_{\mu}$ events at 1 km is then $\sim10^{9}$. The
number of events above the $\tau$ threshold is $\sim10^{8}$.

At a baseline of 1 km, the $\nu_{\mu}$ disappearance and $\nu_{\tau}$
appearance probabilities can be approximated as follows:
\begin{eqnarray*}
P_{\nu_{\mu}\rightarrow\nu_{\mu}}&=&1-4c^{2}_{14}s^{2}_{24}(1-c^{2}_{14}s^{2}_{24})\sin^{2}\Delta_{s}=
1-\sin^{2}2\theta_{s}^{\mu\mu}\sin^{2}\Delta_{s}\\
P_{\nu_{\mu}\rightarrow\nu_{\tau}}&=&4c^{4}_{14}s^{2}_{24}c^{2}_{24}s^{2}_{34}\sin^{2}\Delta_{s}=
\sin^{2}2\theta_{s}^{\mu\tau}\sin^{2}\Delta_{s}
\end{eqnarray*}where c$_{ij}=\cos\theta_{ij}$, s$_{ij}=\sin\theta_{ij}$ and $\Delta_{s}=\Delta m_{s}^{2}L/4E$.

\begin{figure}[tb]
\centering{
\includegraphics[scale=0.7]{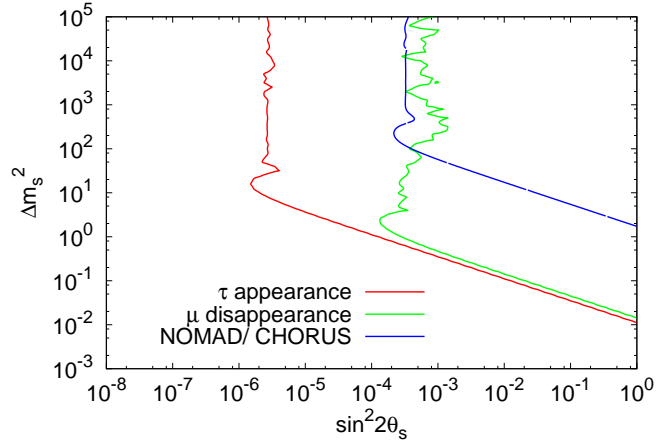}}
\caption{Comparison between the sensitivity reach with 
the discovery channel and that with the disappearance channel.}
\label{Fig:Li}
\end{figure}

In the figure below (Fig.~\ref{Fig:Li}) we demonstrate how the
$\nu_{\tau}$ \textit{appearance} channel (red line) has a more
powerful reach than the $\nu_{\mu}$ \textit{disappearance} channel
(green line), qualitatively comparing the sensitivities of the two
channels to $\sin^{2}2\theta$ and $\Delta m^{2}$ (where
$\sin^{2}2\theta = \sin^{2}2\theta_{s}^{\mu\mu}$ or
$\sin^{2}2\theta_{s}^{\mu\tau}$). We use hypothetically `perfect'
$\nu_{\mu}$ detection ($100\%$ efficiency, zero background, negligible
systematic errors) corresponding to $\sim10^{9}$ charged-current
$\nu_{\mu}$ events, comparing it with that of realistic $\nu_{\tau}$
detection ($10\%$ efficiency, 2 background events, $10\%$ systematic
errors) corresponding to $\sim10^{7}$ events above the $\tau$
threshold. Also shown are the current bounds on
$\sin^{2}2\theta^{\mu\tau}$ and $\Delta m^{2}$, as obtained by the
NOMAD and CHORUS experiments (blue line).

The MINSIS combination of high statistics and good background
rejection produces an impressive physics reach, improving on current
bounds by a factor of $\sim 100$ for both $\sin^{2}2\theta^{\mu\tau}$
and $\Delta m^{2}$ using the flux described above. We find that using
a beam with a peak energy above $\sim$ 12 GeV would produce the best
results, together with maximizing the detector efficiency and
background rejection. Systematic errors and energy resolution have
negligible effects on the experimental sensitivity.

%%%%%%%%%%%%%%%%%%%%%%%%%%%%%%%%%%%%%%%%%%%%%%%%%%%%%%%%%%%%%%%%%%%%%%
\subsection{New physics searches with near detectors at a neutrino factory
    ---   {\it W. Winter}}
%--------------------------------------------------------------------%
\label{winter}

\begin{figure}[t]
\includegraphics[width=0.6\textwidth]{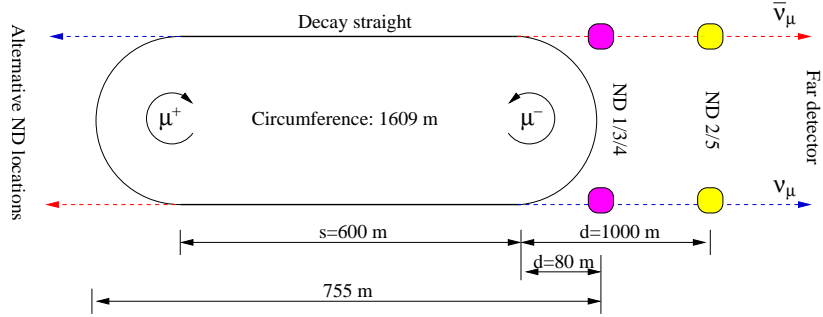}
\caption{\label{fig:ring}Geometry of the muon storage ring and possible
 near detector (ND) locations (not to scale). The baseline $L$ is the
 distance between production point and near detector, \ie, $d \le L \le
 d+s$. Figure taken from \Ref~\cite{Tang:2009na}.}
\end{figure}

Near detectors at a neutrino factory are required for standard
oscillation physics to measure the $\nu_\mu$ and $\bar\nu_\mu$ cross
sections, to monitor the beam (such as by elastic scattering or
inverse muon decay interactions), and to control the backgrounds (such
as the ones from charm production). If the $\mu^+$ and $\mu^-$
circulate in different directions (\cf, \figu{ring}), two near
detectors are required. In this case, flavor identification is
sufficient to measure the inclusive $\nu_\mu$ and $\bar\nu_\mu$ cross
sections. Charge identification is needed for the background
measurements.  As it is demonstrated in \Ref~\cite{Tang:2009na}, the
size, location, and geometry of the near detectors hardly matter for
standard oscillation physics even in extreme cases of possible near
detectors. Because of the high statistics in all energy bins of the
near detectors, the physics potential is generally limited by the
statistics in the far detector(s). Therefore, some characteristics of
the near detectors, such as the location, may be driven by new physics
searches.  However, note that rare interactions used for flux
monitoring, such as inverse muon decays or elastic scattering, may
require large enough detectors. A possible near detector design for a
neutrino factory is, for instance, discussed in
\Ref~\cite{Abe:2007bi}.

Comparing potential new physics searches at a neutrino factory and the
MINSIS superbeam based detector, the two beams have different
characteristics. At the neutrino factory, neutrinos are produced from
muon decays, implying that both $\nu_\mu$ ($\bar\nu_\mu$) and
$\bar\nu_e$ ($\nu_e$) are in the beam for $\mu^-$ ($\mu^+$) stored,
50\% each. The origin of the neutrinos is typically determined by
charge identification of the secondary particle in the detector. For
tau neutrino detection, the origin can be $\bar\nu_e \rightarrow
\bar\nu_\tau$ ($\nu_e \rightarrow \nu_\tau$) or $\nu_\mu \rightarrow
\nu_\tau$ ($\bar\nu_\mu \rightarrow \bar\nu_\tau$) transitions. At the
superbeam, the neutrinos are mainly produced through pion decays. Only
$\nu_\mu$ ($\bar \nu_\mu$) are in the beam for $\pi^+$ ($\pi^-$)
decays, with some contamination from other flavors and polarities. For
tau neutrino detection, only $\nu_\mu \rightarrow \nu_\tau$
($\bar\nu_\mu \rightarrow \bar\nu_\tau$) transitions are accessible
with reasonable sensitivities. However, the absence of a significant
amount of $\bar \nu_e$ ($\nu_e$) in the beam may, depending on the
detector technology, also be an advantage with respect to the
suppression of $\bar \nu_e$ ($\nu_e$) charm induced backgrounds. In
summary, the new physics searches at a neutrino factory and superbeam
may be very complementary if the new physics effect
\begin{itemize}
\item
 is only present in either muon decays or pion decays (such as
 leptonic versus hadronic source NSI)
\item
 requires either low backgrounds (superbeam) or the $\bar\nu_e
 \rightarrow \bar\nu_\tau$ ($\nu_e \rightarrow \nu_\tau$) channel
 (neutrino factory).
\end{itemize}

\begin{figure}[tb]
\begin{center}%\vspace{-0.7cm}
\includegraphics[width=6cm,bb=0 0 720 720]{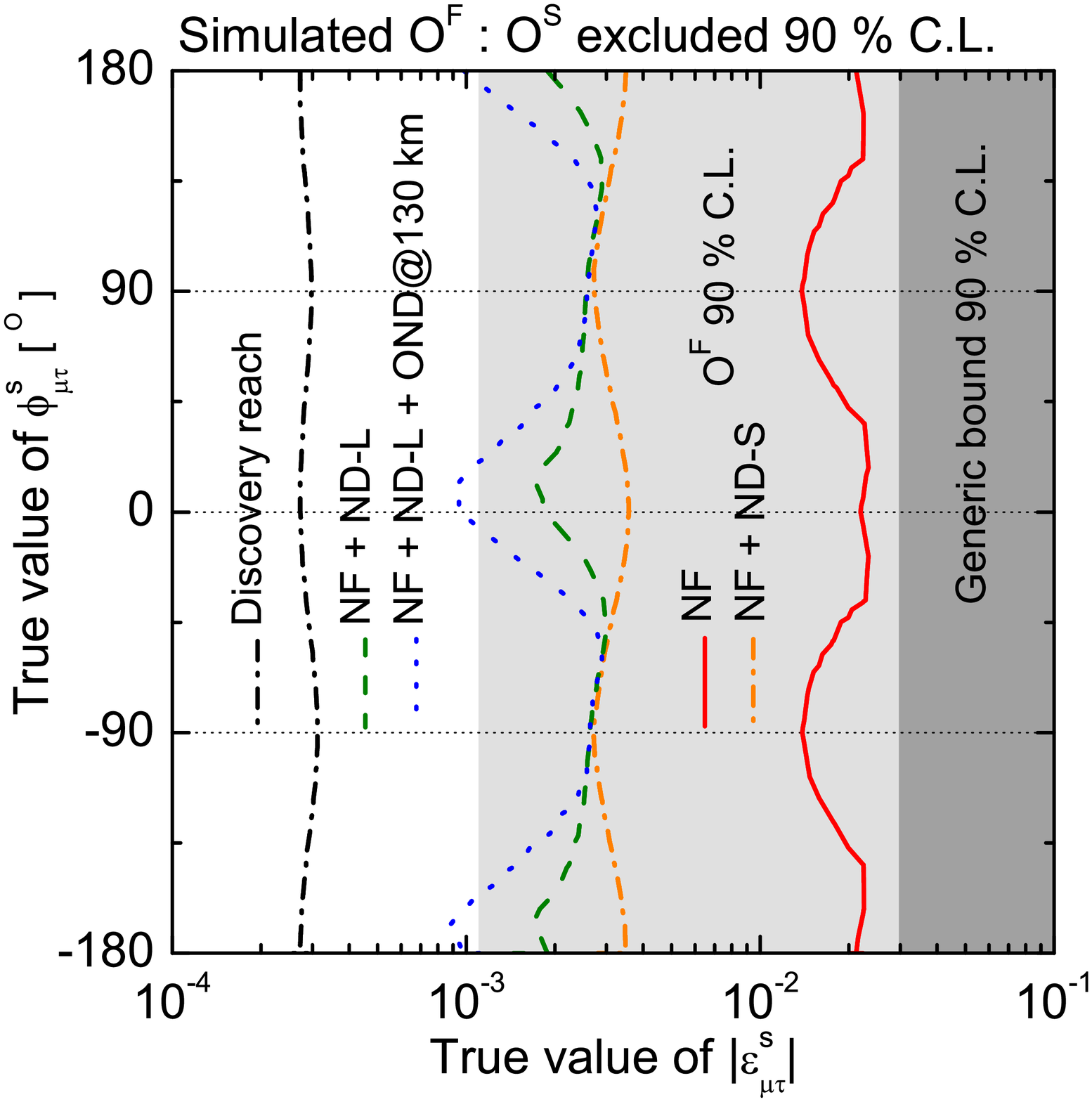}
\includegraphics[width=6cm,bb=0 0 720 720]{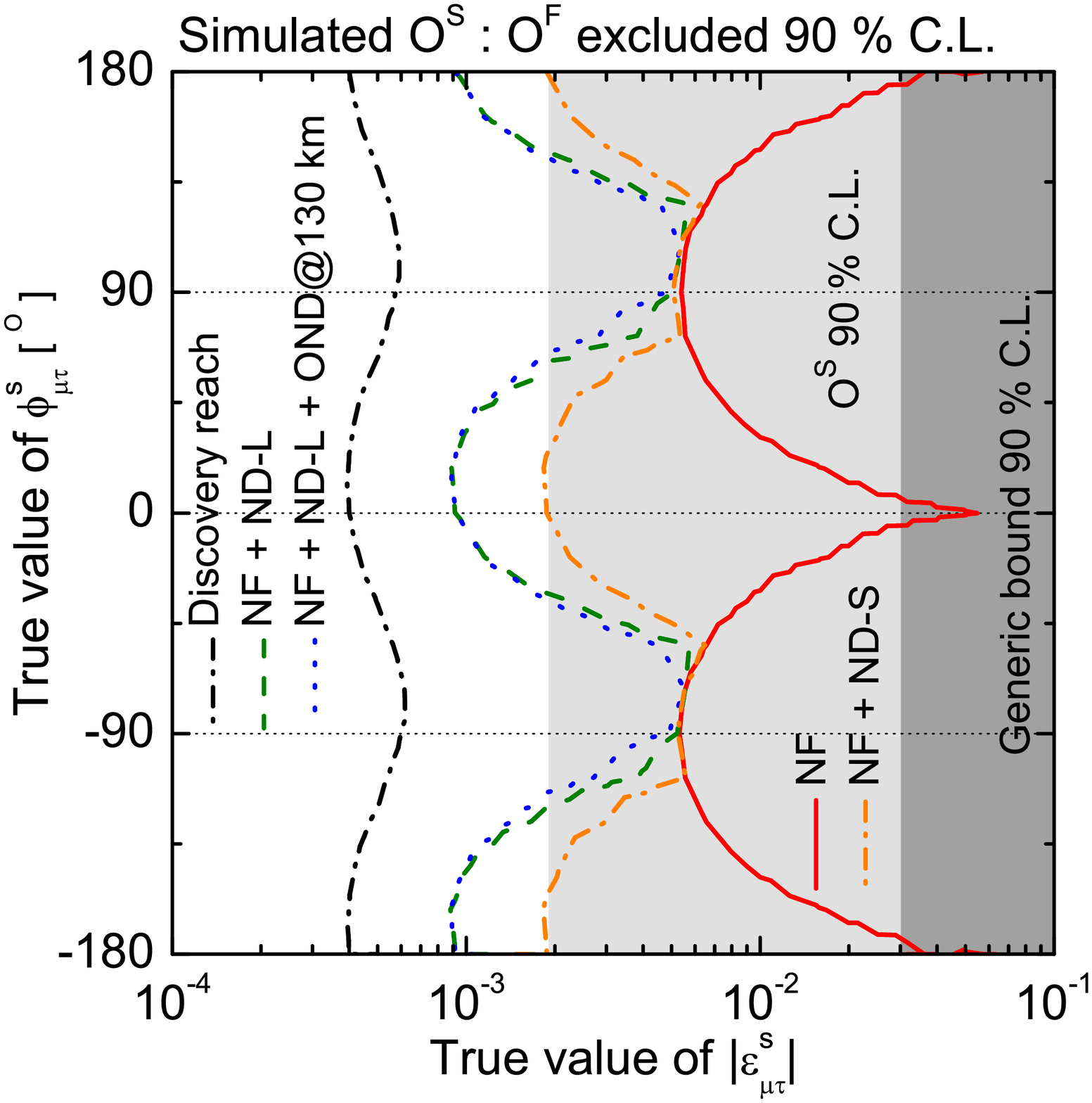}
\vspace{-0.cm}
\caption{\label{fig:muvnsi} %\it 
Regions in the ($|\varepsilon^s_{\mu\tau}|$-$\phi^s_{\mu\tau}$)-plane
where the simulated $\epsilon^s_{\mu\tau}$ induced by one type of
operator can be uniquely established, \ie, the other type of operator
is excluded at the 90\% C.L.  (regions on the right-hand side of the
curves). Left panel: the simulated $\varepsilon^s_{\mu\tau}$ is
induced by ${\cal O^F}$ (non-unitarity) and fitted with ${\cal O^S}$
(NSI from $d=6$ operator). Right panel: the simulated
$\varepsilon^s_{\mu\tau}$ is induced by ${\cal O^S}$ and fitted with
${\cal O^F}$. The different curves corresponds to the IDS-NF baseline
setup (NF), an additional (small) silicon vertex-sized near detector
(ND-S), an additional OPERA-like near detector at 1~km (ND-L), and an
additional OPERA-like near detector at 130~km (OND@130~km).  In both
panel, the discovery reach is also displayed. Figure taken from
\Ref~\cite{Meloni:2009cg}. }
\end{center}
\end{figure}

Let us illustrate this complementarity with one example from
\Ref~\cite{Meloni:2009cg}. If new physics comes from heavy mediators,
which are integrated out above the electroweak symmetry breaking
scale, the lowest order contributions (apart from neutrino mass) to
the Standard Model come from effective $d=6$ operators. Heavy neutral
fermions lead to an addition to the kinetic energy of the neutrinos,
also known as minimal flavor violation, which implies a non-unitary
(NU) mixing matrix after the re-diagonalizing and re-normalization
of the kinetic terms of the neutrinos. Heavy bosons, such as singly
charged scalar SU(2) singlets, on the other hand, lead to non-standard
interactions (NSI) at tree level.  Therefore, distinguishing between
NU and NSI may be interpreted as distinguishing between fermions and
bosons as heavy mediators, at least to leading order at $d=6$ and tree
level.  We therefore refer to these effects at $d=6$ as
$\mathcal{O^F}$ and $\mathcal{O^S}$, respectively.  At a neutrino
factory, the phenomenology of $\mathcal{O^F}$ and $\mathcal{O^S}$ is,
however, very similar. Both effects can be parametrized in form of
NSI. For $\mathcal{O^F}$, particular correlations among source,
propagation, and detection effects are
present~\cite{FernandezMartinez:2007ms,Antusch:2008tz}. For leptonic
$\mathcal{O^S}$, similar correlations are obtained for operators
without charged lepton flavor
violation~\cite{Gavela:2008ra}. Consider, for instance,
\begin{eqnarray}
\varepsilon^m_{\mu \tau } = & - (\varepsilon^s_{\mu \tau})^* & \quad
\text{(NSI)} \, ,
\label{equ:corr1}\\
\varepsilon^m_{\mu \tau} = & - \varepsilon^s_{\mu \tau} & \quad
\text{(NU)} \, .
\label{equ:corr2}
\end{eqnarray}
In this case, the two effects can be mostly distinguished by the
absence of detection effects in the case of ${\cal
O^S}$. Alternatively, one could use a superbeam-based source, where
\equ{corr1} does not hold because of the neutrino production by pion
decays.  We illustrate the identification of the class of effect in
\figu{muvnsi}. Obviously, the effects can be hardly distinguished with
a neutrino factory alone beyond the current bounds.  However, the
discovery reach for the non-standard effects clearly exceeds the
current bounds.  The MINSIS detector could disentangle the two effects
in the region between the bounds and the discovery reach if it had a
sensitivity significantly exceeding $10^{-3}$ in $|\epsilon_{\mu
\tau}^s|$.

%\providecommand{\bysame}{\leavevmode\hbox to3em{\hrulefill}\thinspace}

%%%%%%%%%%%%%%%%%%%%%%%%%%%%%%%%%%%%%%%%%%%%%%%%%%%%%%%%%%%%%%%%%%%%%%
\subsection{Very short baseline electron neutrino disappearance
    --- {\it M. Laveder}}
%--------------------------------------------------------------------%
\label{lavender}
In Ref.\cite{Giunti:2009zz} possible indications of
Very-Short-BaseLine (VSBL) electron neutrino disappearance into
sterile neutrinos in MiniBooNE neutrino data and Gallium radioactive
source experiments have been considered.  The compatibility of such a
disappearance with reactor and MiniBooNE antineutrino data has been
discussed.  A tension between neutrino and antineutrino data has been
found, which could be due to: 1) statistical fluctuations; 2)
underestimate of systematic uncertainties; 3) exclusion of our
hypothesis of VSBL $\nu_e$ disappearance; 4) a violation of CPT
symmetry.  Considering the first possibility, the results of a
combined fit of all data have been presented, which indicate that
$P_{ee} < 1$ with 97.04\% \, CL.  The possibility of CPT violation has
been considered, which leads to the best-fit value $A_{ee}^{CPT,bf} =
-0.17 \pm 0.05$ for the asymmetry of the $\nu_e$ and $\bar{\nu_e}$
survival probabilities and $ A_{ee}^{CPT} < 0 $ at 99.7\% \,CL.
%%%%%%%%%%%%%%%%%%%%%%%%%%%%%%%%%%%%%%%%%%%%%%%%%%%%%%%%%%%%%%%%%%%%%%% 
%%%%%%%%%%%%%%%%%%%%%%%%%%%%%%%
This result translates in an oscillation amplitude $ \sin^2  
2\theta_{es} = 0.34 \pm 0.10 $
for the active-sterile electron neutrino oscillation to be searched  
for in coming SBL experiments
like the MINSIS proposal.
%%%%%%%%%%%%%%%%%%%%%%%%%%%%%%%%%%%%%%%%%%%%%%%%%%%%%%%%%%%%%%%%%%%%%%% 
%%%%%%%%%%%%%%%%%%%%%%%%%%%%%%%
In Ref. \cite{Giunti:2009en}
short-baseline and very-short-baseline $\nu_e$ disappearance at a  
neutrino factory have been studied.
Geometric effects, such as from averaging over the decay straights,
and the uncertainties of the cross sections were taken into account.
An approach similar to reactor experiments with two detectors were  
followed:
two sets of near detectors at different distances were used to cancel  
systematics.
It was demonstrated that such a setup is very robust with respect to  
systematics,
and can have excellent sensitivities to the effective mixing angle and  
squared-mass splitting.
In addition, the possibility of CPT violation
can be tested (depending on the parameters) up to a 0.1\% level.

%%%%%%%%%%%%%%%%%%%%%%%%%%%%%%%%%%%%%%%%%%%%%%%%%%%%%%%%%%%%%%%%%%%%%%
\subsection{Tau detection using the kinematic and impact parameter techniques
    ---   {\it F.~J.~P. Soler}}
%--------------------------------------------------------------------%
\label{soler}
\hspace*{-0.4cm}
{\bf Tau detection techniques}

 Currently there are three possible techniques for detecting taus:
 \begin{itemize}
 \item Direct observation of the decay kink of the tau with the use 
 of emulsion technology, like in OPERA \cite{Acquafredda:2009zz} or CHORUS
 \cite{Eskut:2007rn};
 \item The identification of taus through the kinematic analysis of 
 the tau decay, like that used by NOMAD \cite{Astier:2001yj};
 \item The reconstruction of taus from an impact parameter signature 
 with a dedicated silicon vertex detector, as in the NAUSICAA 
 proposal \cite{GomezCadenas:1995ij}, 
 prototyped by NOMAD-STAR \cite{Barichello:2003gu}.
 \end{itemize}

\hspace*{-0.4cm}
 {\bf NOMAD}

 NOMAD was a $\nu_\mu\rightarrow \nu_\tau$ neutrino oscillation 
 experiment at the CERN SPS between 1994-1998 \cite{Astier:2001yj}.
 The main aim was to search for the appearance of $\nu_\tau$ in a 
 predominantly $\nu_\mu$ beam. A total of  $1.35\times 10^6~\nu_\mu
 $~charged current (CC) events were recorded in NOMAD for 
 $5\times10^{19}$ protons on target (pot). NOMAD used the kinematic 
 technique, where the visible products from the tau decay are 
 measured, and kinematically separated from background by exploiting 
 that taus decay emitting one or two neutrinos (which are not 
 observed) thereby producing events with larger missing transverse 
 momentum ($p_t$) than normal $\nu_\mu$ CC events.
 NOMAD was sensitive to 82.4\% of the branching fraction of the taus. 
 The analysis exploited a set of likelihood functions that 
 parametrized the missing $p_t$ and isolation of the tau candidates. 
 The final sensitivity achieved was $P(\nu_\mu\rightarrow \nu_
 \tau)<1.63\times 10^{-4}$ at 90\% confidence level. In the NOMAD 
 analysis, a number of kinematic regions of the tau decay phase space 
 were exploited to optimize the sensitivity. However, about half the 
 sensitivity came from low background bins, which could in principle 
 be exploited further, in a higher statistics experiment. Therefore, 
 there is room for improvement of the kinematic technique in a higher 
 statistics experiment, like the MINSIS experiment being proposed at 
 Fermilab. A liquid argon experiment could, in principle, carry out a 
 similar analysis and further exploit this analysis technique at 
 MINSIS. One would need to take into account the intrinsic tau 
 contamination of the neutrino beam from $D_s$ decay, which was 
 estimated to be $3.5\times 10^{-6}$ in the CHORUS-NOMAD beam (450 
 GeV), going down to $9.6\times 10^{-8}$ at the Main Injector at 
 Fermilab (120 GeV) \cite{VandeVyver:1996qc,GonzalezGarcia:1996ri}.

\vspace{0.2cm}
\hspace*{-0.4cm}
 {\bf Impact parameter detection of taus}

 The impact parameter technique for detection of taus was first 
 proposed by Gomez Cadenas et al. in a proposal called NAUSICAA 
 \cite{GomezCadenas:1995ij}. A  silicon vertex detector with a $B_4 C$ target 
 was proposed as an ideal medium to identify taus. Standard $\nu_\mu$ 
 CC interactions have an impact parameter resolution of 28~$\mu$m, 
 while tau decays have an impact parameter resolution of 62~$\mu$m. 
 By performing a cut on the impact parameter significance ($
 \sigma_{IP}/IP$) one can separate  one prong decays of the tau from 
 the background. For three prong decays of the tau, a double vertex 
 signature is used to separate signal from background. The total net 
 efficiency of the tau signal in NAUSICAA was found to be 12\%. With 
 this efficiency, one could have a sensitivity of  $P_{\mu\tau}< 
 3\times 10^{-6}$ at 90\% C.L. on the $\mu-\tau$ conversion 
 probability.

 Another idea proposed in 1996 was to use a hybrid detector emulsion-
 silicon tracking to improve the tau detection efficiency 
 \cite{GomezCadenas:1996jx}. A Letter of Intent (called TOSCA) was submitted to the 
 CERN SPSC in 1997 \cite{Ayan:1997pm} with a detector based around this 
 idea.  Tau detection efficiencies of 42\%, 10.6\% and 27\% were 
 determined for the muon, electron and one charged hadron decays of 
 the tau, yielding a net probability of $P_{\mu\tau}< 0.75\times 
 10^{-5}$ for the CERN SPS beam at 350 GeV. A program of R\&D, 
 called NOMAD-STAR, including a 50~kg prototype silicon-$B_4 C$ 
 target, operated between 1997-1998 in the NOMAD beam
       \cite{Barichello:2003gu}. 
It was able to demonstrate an impact parameter resolution of 
 $33~\mu$m and a double vertex resolution of $18~\mu$m, which were 
 the expected parameters to achieve the $\nu_\mu-\nu_\tau$ 
 sensitivity of NAUSICAA and TOSCA. About 45 charm events were 
 detected with NOMAD-STAR over the duration of the run \cite{Ellis:2003vq}.

\vspace{0.2cm}
\hspace*{-0.4cm}
 {\bf Tau detection at a near detector of a neutrino factory}

 A near detector at a neutrino factory needs to  measure the charm 
 cross-section to validate the size of the charm background in the 
 far detector, since this is the main background to the wrong-sign 
 muon signature. The charm cross-section and branching fractions are 
 poorly known, especially close to threshold, so this detector would 
 need to be able to detect charm particles. Since tau events have a 
 similar signature to charm events, any detector that can measure 
 charm should be able to measure taus as well. A semiconductor vertex 
 detector is the only viable option in a high intensity environment 
($ \sim 10^9~\nu_\mu$ CC events per year in a detector of mass 1 ton), 
 for charm detection, since a liquid argon detector would not be fast 
 enough to cope with the rate and emulsion would perish in this 
 environment.

 Assuming 12\% efficiency from the NAUSICAA proposal, and assuming
 that charm production is about 4\% of the $\nu_\mu$ CC rate between
 10 and 30 GeV (CHORUS measured $6.4\pm 1.0\%$ at 27 GeV)
 \cite{Lellis:2004yn}, would imply a signal of $1.2\times 10^8$ tau
 events (with $P_{\mu\tau}=3D100\%$ conversion rate) and $4\times
 10^7$ charm events. Charm events from anti-neutrinos (for example
 $\overline{\nu}_e$) mimic the potential signal. The identification of
 the positron can reduce the background, but electron and positron
 identification normally has a lower efficiency than muon
 identification. It is very important to have a light detector (ie, a
 scintillating fibre tracker) behind the vertex detector inside a
 magnetic field to identify the positron with high efficiency (in the
 best scenarios $\sim 80\%$ would be the maximum achievable). A
 further way to separate the charm background from signal is to use
 the kinematic techniques of NOMAD. Assuming the NOMAD net efficiency
 yields a $P_{\mu\tau}< 2\times 10^{-6}$, which is not better than
 what the MINSIS detector could achieve at Fermilab.

\vspace{0.2cm}
\hspace*{-0.4cm}
 {\bf Summary}

 So, in summary, a MINSIS detector based on the emulsion cloud 
 chamber technique (like OPERA) could potentially achieve a $P_{\mu
 \tau}$ sensitivity of $10^{-6}$. However, there is also the 
 potential for a liquid argon detector to perform a tau search using 
 the kinematic technique as in NOMAD. While it is likely that the 
 sensitivity might not be as low as the OPERA like detector, it could 
 serve as very useful R\&D for liquid argon, and provide a useful 
 physics outcome.

 A TOSCA like detector, with silicon and emulsion, could also be done 
 for MINSIS. However, adding silicon complicates things and the 
 sensitivity gain is not obvious any more, since scanning technology 
 has advanced so much that one could potentially scan all the 
 emulsion obviating the need for the silicon detector.

 A silicon target only (as in NAUSICAA) has less efficiency but does 
 not rely on emulsion. The advantage of this approach would be that 
 the analysis could be performed faster than with emulsion and, in 
 principle could achieve a limit of about $3\times 10^{-6}$. At a 
 neutrino factory near detector one can also measure charm and taus 
 using a silicon tracker. Neither emulsion nor liquid argon would be 
 suitable at a neutrino factory near detector since the event rate is 
 too high. However, the background at a neutrino factory is higher 
 than at a conventional neutrino beam from pion decay, since there is 
 a charm background from anti-neutrinos. This background could, in 
 principle, be reduced with a combination of the impact parameter and 
 kinematic approach for tau detection, but the sensitivity  of 
 $2\times 10^{-6}$ would not be any better than what could be 
 achieved at MINSIS.

%%%%%%%%%%%%%%%%%%%%%%%%%%%%%%%%%%%%%%%%%%%%%%%%%%%%%%%%%%%%%%%%%%%%%%
\pagebreak

\section{Workshop Summary}

The MINSIS idea aims at a measurement of $\tau$ appearance events in a
near position close to the NuMI beam target with a minimum senstivity of
$10^{-6}$. 
The talks ~Secs.~\ref{para}, ~\ref{migliozzi}, and~\ref{soler} discussed  
the challenges and requirements for  achieving  
such sensitivity with a ``standard'' emulsion detector as well as
with possible variations using a detector with a pure  silicon target 
or a silicon-emulsion combination.

The theoretical motivation was the core of the discussions in the workshop.
On one hand it is clear that the discovery of $\tau$ appearance 
at short distances at rates above  $10^{6}$ would be an exciting evidence for 
non-standard physics. 
On the other hand resides the challenge of finding theoretically 
motivated forms of new physics which could give signals at this level.

The first consideration is the requirement on the sensitivity. 
For the MINSIS
baseline and energy one obtains for standard $\nu_\mu\to\nu_\tau$
oscillations induced by the ``atmospheric'' mass difference
\begin{equation}
P^{\rm atm} (\nu_\mu \rightarrow \nu_\tau) = \sin^22\theta_{23} \sin^2(\Delta
m^2_{31}L/4E) \approx 10^{-7}\left(\frac{L}{1 ~{\rm km}} \cdot \frac{10 ~\rm{GeV}}{E}\right)^2. \nonumber
\end{equation}
Hence, the atmospheric oscillations represent an irreducible background, at the level of $10^{-7}$, for the typical energy of the NUMI beam for $\tau$ appearance of 10 GeV or higher. 
This sets the natural limit for the sensitivity. 
%New physics/atmospheric oscillation interference effects have to be taken into account in that regime.

We discussed  two possible forms of new physics  which generically can 
induce $\tau$ appearance at short distance:
\begin{itemize}
\item
{\bf Sterile neutrinos  with masses in the range $m_s
\gtrsim 1$~eV}. Such states are a theoretical possibility.  
The open question is the theoretical motivations 
for sterile neutrinos in that mass range. 
From the purely phenomenological perspective current bounds in the relevant 
parameter range are of order $10^{-4}$ and hence, MINSIS could explore a 
new parameter space of about 2 to 3 orders of magnitude.

\item
{\bf Non-standard neutrino interactions (NSI) and Non Unitarity (NU)
in the leptonic mixing matrix}. Theoretically these two forms of new physics
can be of very different origin but phenomenologically they 
are closely related to each other as stressed in several of
the  talks in this workshop.  A discovery of NSI or NU would be an important 
step towards our understanding of the mechanism of neutrino mass
generation. This would be exciting complementary information 
to data from oscillations, charged lepton flavour violation,
neutrinoless double beta decay, and the LHC.

The challenge is the level at which they can be 
expected.  Theoretically one does not expect a signal
above the $10^{-6}$ level as long as the low energy operators respect  
$SU(2)$. In brief, once an effective operator which induces NSI/NU is 
written down, the same operator (or operators derived from gauge invariance 
and similar arguments) will induce charged lepton flavour violation, which is
severely bounded. Taking into account these bounds one arrives at the
conclusion that, under rather generic assumptions, neutrino
NSI/NU are not expected to be detectable above the level of
$10^{-7}-10^{-6}$ (and often are even much smaller than this). 
The caveat to this argument is that we have no proof that it cannot be 
avoided in specifically constructed models, although an explicit example 
for such a model is still missing.
\end{itemize}

In brief, the $10^{-6}$ sensitivity is the minimum requirement for an
unambiguous identification of the new physics signal. To find ``well
motivated'' theoretical frameworks that predict signals at this level
while complying with all the existing bounds is possible but challenging.
Conversely, if a signal was observed at this level, we would be
confronted with the exciting news of an unexpected form of new
physics.

\section*{Acknowledgments}The participants of this meeting would especially like to thank Prof. Belen Gavela of UAM for her wonderful hospitality during this workshop.
%%%%%%%%%%%%%%%%%%%%%%%%%%%%%%%%%%%%%%%%%%%%%%%%%%%%%%%%%%%%%%%%%%%%%
\pagebreak
\bibliography{./MINSIS-Madrid}
\bibliographystyle{apsrev}

\end{document}